# Superconductivity in Al-based high-entropy alloys TiHfNbTaAl and TaNbHfZrAl

Junjin Huang[a,#], Wenbo Sun[a,#], Longfu Li[a], Shuangyue Wang[a], Jingjun Qin[a], Rui Chen[a], Zaichen Xiang[a], Yucheng Li[a], Lingyong Zeng [a,*], Huixia Luo [a,*]

[a] School of Materials Science and Engineering, State Key Laboratory of Optoelectronic Materials and Technologies, Guang-dong Provincial Key Laboratory of Magnetoelectric Physics and Devices, Key Lab of Polymer Composite & Functional Materials, Sun Yat-sen University, Guangzhou 510006, China

# These authors contributed equally to this work

*Corresponding author:

Dr. Lingyong Zeng (zengly25@mail3.sysu.edu.cn)

Prof. Huixia Luo (luohx7@mail.sysu.edu.cn)

**Abstract**

Since the first report of a high-entropy alloy (HEA) superconductor in 2014, HEAs have continued to captivate the interest of superconducting researchers. Owing to the significant degree of disorder inherent in these systems, they serve as exemplary models for examining the properties of materials that exist in states intermediate between crystalline and amorphous structures. Here we present the superconductivity properties and crystal structure of TaNbHfZrAl and TiHfNbTaAl HEAs, which both have the VEC of 4.2 and body-centered cubic (BCC) structure. Through resistivity, magnetic, and specific heat measurements, we prove that both samples are the bulk type-II superconductors with a critical temperature $T_c$ of 5.5 K for TaNbHfZrAl and $T_c$ of 3.2 K for TiHfNbTaAl. The $T_c$ of HEA superconductors is influenced by the VEC and the element composition. And the incorporation of Al in high disorder HEA superconductors causes a more crystallinelike $T_c$ dependence. We also found that the $\Delta C_{el}/\gamma T_c^{mid}$ (2.31) and $2\Delta_0/K_B T_c$ (4.48) of the TaNbHfZrAl sample are close to the known strongly coupled systems, suggesting that it is a strongly coupled superconductor. The TaNbHfZrAl HEA system serves as a novel material platform for exploring strongly coupled phenomena.

## I. Introduction

In recent years, high-entropy alloys (HEAs), made up of at least five different metallic elements, each ranging from 5% to 35%, have been widely studied [1-3]. The researchers found that HEAs have extremely stable structures, such as the hexagonal-closed packed (HCP), body-centered cubic (BCC), and face-centered cubic (FCC) [3]. The reason why HEAs have stable structures is that the high configurational entropy resulting in the random occupation of each element in the lattice site dominates the free energy. Many recent observations have proven that HEAs display superior mechanical, chemical/physical properties compared with conventional alloys, which are potentially employed under extreme conditions [4-8].

The initial HEA superconductor, $Ta_{34}Nb_{33}Hf_8Zr_{14}Ti_{11}$, was announced in 2014. It adopts a BCC structure with a critical temperature ($T_c$) of 7.3 K, a lower critical field of 32 mT, and an upper critical field of 8.2 T. The electronic density of states (DOSs) at the Fermi level has an energy gap of 2.2 meV, characteristic of a Bardeen-Cooper-Schrieffer (BCS) phonon-mediated superconductor [9]. Since then, HEA superconductors have been studied increasingly [2]. In some HEA superconductors, some exotic superconductivities have been observed, including maintaining zero resistance at high pressure [10-11], large upper critical field [12-14], higher critical current density [6, 15], strong-coupling behavior of the *s*-wave superconductors [16-18], and type-II Dirac points in the band structures [7, 19]. However, for most HEA superconductors, the $T_{cs}$ are lower than 10 K, and the average valence electron count (VEC) and Nb element content have an important influence on $T_c$ [2, 20]. Recently, TaNbHfZr medium-entropy alloy (MEA) was reported to show dome-shaped superconductivity under pressure, and at 71.6 GPa, $T_c$ is 15.3 K, which is the highest recorded $T_c$ of HEA superconductors [21]. Additionally, TiHfNbTa MEA was found to show a bulk $T_c$ of around 6.75 K with extremely strong coupling ($\Delta C_{el}/\gamma T_c^{mid}$=2.88, $2\Delta_0/K_BT_c$=5.02) [17]. Furthermore, Chen et. al. found that with the increase of aluminum content in HEAs, the hardness of the material can be improved, and the material changes from a plastic material to a medium-low temperature brittle material [22]. Through first-principle calculation, Levente et al. found that the increase of doped aluminum content can reduce the twin property of CrMnFeCoNi HEA, and reduce the unstable laminar faults and unstable twin faults [23]. Therefore, we generally believe that doping with aluminum can improve the mechanical properties of HEA. However, there are still very few studies on HEA superconductors with aluminum (Al) doping. Therefore, we doped the Al element into the TaNbHfZr and TiHfNbTa MEAs and synthesized TaNbHfZrAl and TiHfNbTaAl HEAs to understand the effect of VEC change on superconductivity.

In this study, we present the superconducting properties and crystal structure of TaNbHfZrAl and TiHfNbTaAl HEAs, both of which have a VEC of 4.2 and body-centered cubic (BCC) structure. Through specific heat, resistivity, and magnetic susceptibility measurements, we prove that both samples are the bulk type-II superconductors with a $T_c$ of 5.5 K for the TaNbHfZrAl and a $T_c$ of 3.2 K for the TiHfNbTaAl sample. Adding Al element in TiHfNbTa and TaNbHfZr leads to decrease of the $T_c$ with decreasing VEC. We also found that the $\Delta C_{el}/\gamma T_c^{mid}$ (2.31) and $2\Delta_0/K_B T_c$ (4.48) of the TaNbHfZrAl sample are close to the known strongly coupled systems, proving that it is a strongly coupled superconductor.

## II. Materials and Methods

The arc-melting method was used to prepare polycrystalline samples of TaNbHfZrAl and TiHfNbTaAl. The piece was formed by pressing stoichiometric mixtures of high-purity Hf, Zr, Nb, Ta, Ti, and Al powders, which were then melted in the arc furnace. The melting process is under a high-purity Ar atmosphere, followed by rapid cooling on the water-cooled hearth. To guarantee phase uniformity, we flipped and remelted the sample three times. Long heating treatments at high current were avoided, and the weight loss during melting was found to be about 1.5% for both TiHfNbTaAl and TaNbHfZrAl samples. The crystal structure of the resulting HEAs was determined using powder X-ray diffraction (PXRD) through the MiniFlex of Rigaku. The PXRD data were determined from 30$^{o}$ to 90$^{o}$ with a 1 $^{o}$/min scan speed at room temperature. Chemical compositions were determined using the scanning electron microscope (SEM), backscattered electron micrograph (BSEM), and energy dispersive

X-ray spectroscopy (EDX). Besides, the magnetization, resistivity, and heat capacity measurements were measured by a Physical Property Measurement System (PPMS, Quantum Design). The resistivity was measured with the standard four-probe configuration by attaching platinum electrondes with silver paste on the sample. DC magnetization was measured using the vibrating sample magnetometer (VSM) option of the PPMS Dynacool system. Specific heat measurements were performed with the Quantum Design heat-capacity option, using a relaxation technique.

**III. Results and Discussion**

Figure 1(a) and Figure 1(b) display the PXRD patterns for the TaNbHfZrAl and TiHfNbTaAl HEAs. It can be seen that both samples are pure single-phase. Each observed PXRD peak is marked with its corresponding Miller indices. The refined XRD profile indicated that both HEAs show a $Im\bar{3}m$ space group in BCC structure. By analyzing the diffraction data, the lattice parameter was acquired to be $a$ = 3.3717(7) Å for TaNbHfZrAl HEA and $a$ = 3.3266(7) Å for TiHfNbTaAl HEA. Besides, we can discover the mild broadening of the signals of peaks, attributable to a high degree of disorder of the atoms. Through using XRD patterns, we obtained that the crystallite size and microstrain of TaNbHfZrAl are 10.3 nm and 0.92%, and the crystallite size and microstrain of TiHfNbTaAl are 10.7 nm and 0.98%. The atomic size difference parameter δ is a measure of the degree of the atomic size difference among the constituent elements, which is calculated by $\delta = 100 \times \sqrt{\sum_{i=1}^{n} ci(1-\frac{r_i}{\bar{r}})^2}$, where $r_i$ is the atomic radius of element $i$ and $\bar{r}$ is the composition-weighted average atomic radius. Through calculating the δ, we gained the δ of TaNbHfZrAl is 4.76 and the δ of TiHfNbTaAl is 3.74. For comparison, the δ values of TaNbHfZr and TiHfNbTa samples are 4.43 and 3.68, respectively. The introduction of Al element significantly increased the atomic mismatch. In general, the greater the entropy, the more atoms there are, which may lead to higher atomic mismatches. Figure 1(c) reveals the crystal structure of TaNbHfZrAl and TiHfNbTaAl samples, in which the Wyckoff position is randomly occupied by the metal atom, leading to site mixing and disorder. Figure S1 shows the SEM images for the TiHfNbTaAl and TaNbHfZrAl HEAs. From the elemental mappings (see Figure S1(c)), we can see that these two HEAs are microscopically homogeneous. The EDX results give the actual ratio ($Ti_{0.23}Hf_{0.22}Nb_{0.21}Ta_{0.20}Al_{0.14}$ and $Ta_{0.21}Nb_{0.21}Hf_{0.23}Zr_{0.20}Al_{0.15}$) of these two HEAs. The actual content of the Al element is relatively low, which may be because the Al element has a lower boiling point, resulting in a small loss during the melting process.

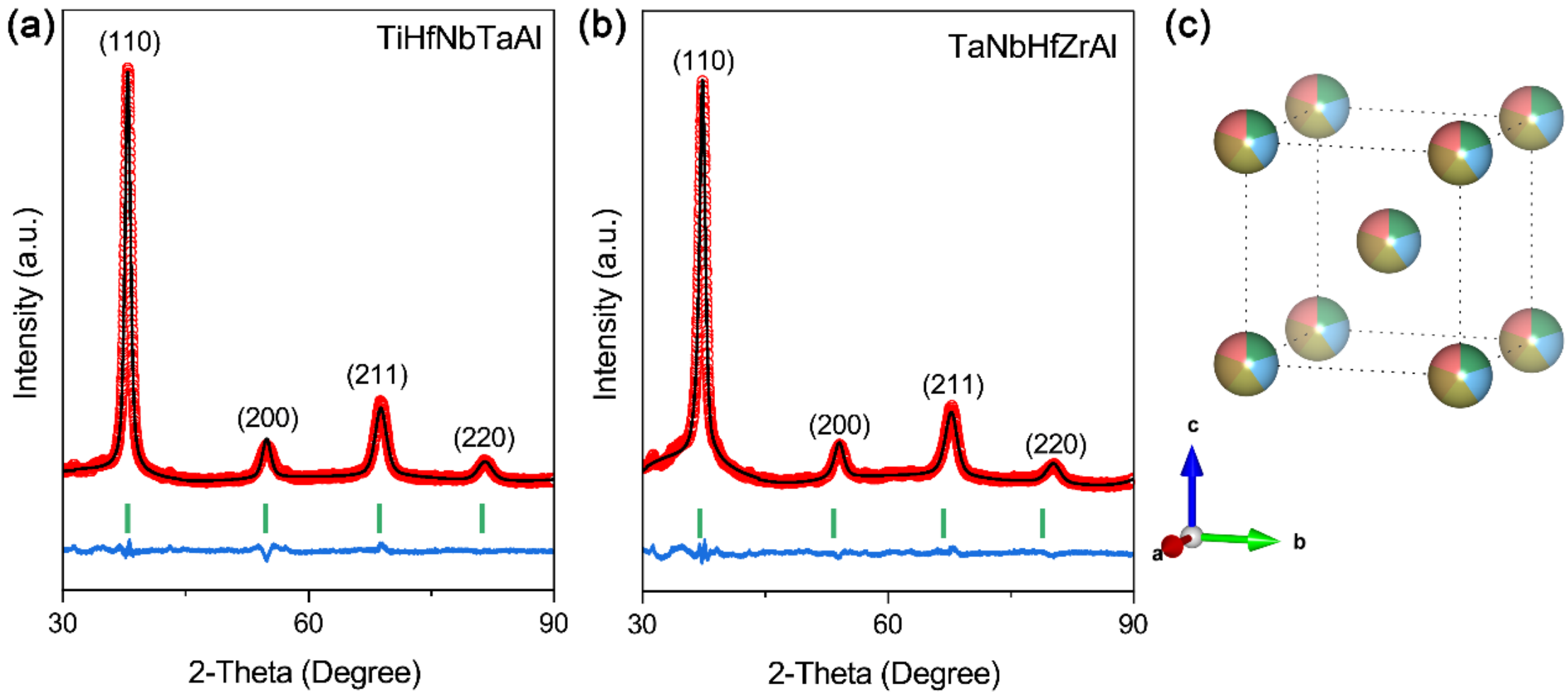

**Figure 1.** (a) Rietveld refinement pattern for the TiHfNbTaAl sample. (b) Rietveld refinement pattern for the TaNbHfZrAl sample. (c) Crystal structure of BCC-type TiHfNbTaAl and TaNbHfZrAl.

Figure 2(a) displays resistivity data of the TiHfNbTaAl and TaNbHfZrAl from 300 to 1.8 K under the zero field. The resistivity of these Al-based HEAs decreases with temperature, exhibiting a poor metallic behavior in the normal state. As the temperature gets lower, a sharp drop of resistivity arises at $T_c$ = 3.1 K of TiHfNbTaAl and $T_c$ = 5.5 K of TaNbHfZrAl, implying the appearance of the superconducting state. We defined the temperature at the midpoint of the superconducting transition as the $T_c$. The detailed superconducting transition for TiHfNbTaAl HEA is shown in Figure 2(b) and for TaNbHfZrAl HEA in Figure 2(c). To study the upper critical field $\mu_0H_{c2}$ of these two Al-based HEAs, we measured the temperature-dependent resistivity under several magnetic fields. The dependence of $T_c$ on the applied field was displayed in Figure 2(d) for TiHfNbTaAl HEA and Figure 2(e) for TaNbHfZrAl HEA, where the field is increased from 0 to 2 T for TiHfNbTaAl and from 0 to 7 T for TaNbHfZrAl. Figure 2(f) displays the $\mu_0H_{c2}$ plotted as a function of the estimated $T_c$ values and fitted with the Ginzburg-Landau (GL) equation for TiHfNbTaAl and TaNbHfZrAl. The GL formula is given as $\mu_0H_{c2}(T) = \mu_0H_{c2}(0) \times \frac{1-(T/T_c)^2}{1+(T/T_c)^2}$. The 0 K upper critical field $\mu_0H_{c2}(0)$ was estimated as 3.7(7) T for TiHfNbTaAl and 8.5(3) T for TaNbHfZrAl. The Pauli-limit field, as explained by the BCS theory, is defined as $\mu_0H_{c2}^P(0) = 1.85 \times T_c$, and then $\mu_0H_{c2}^P(0)$ is calculated to be 5.7(4) T for TiHfNbTaAl, 9.9(9) T for TaNbHfZrAl.

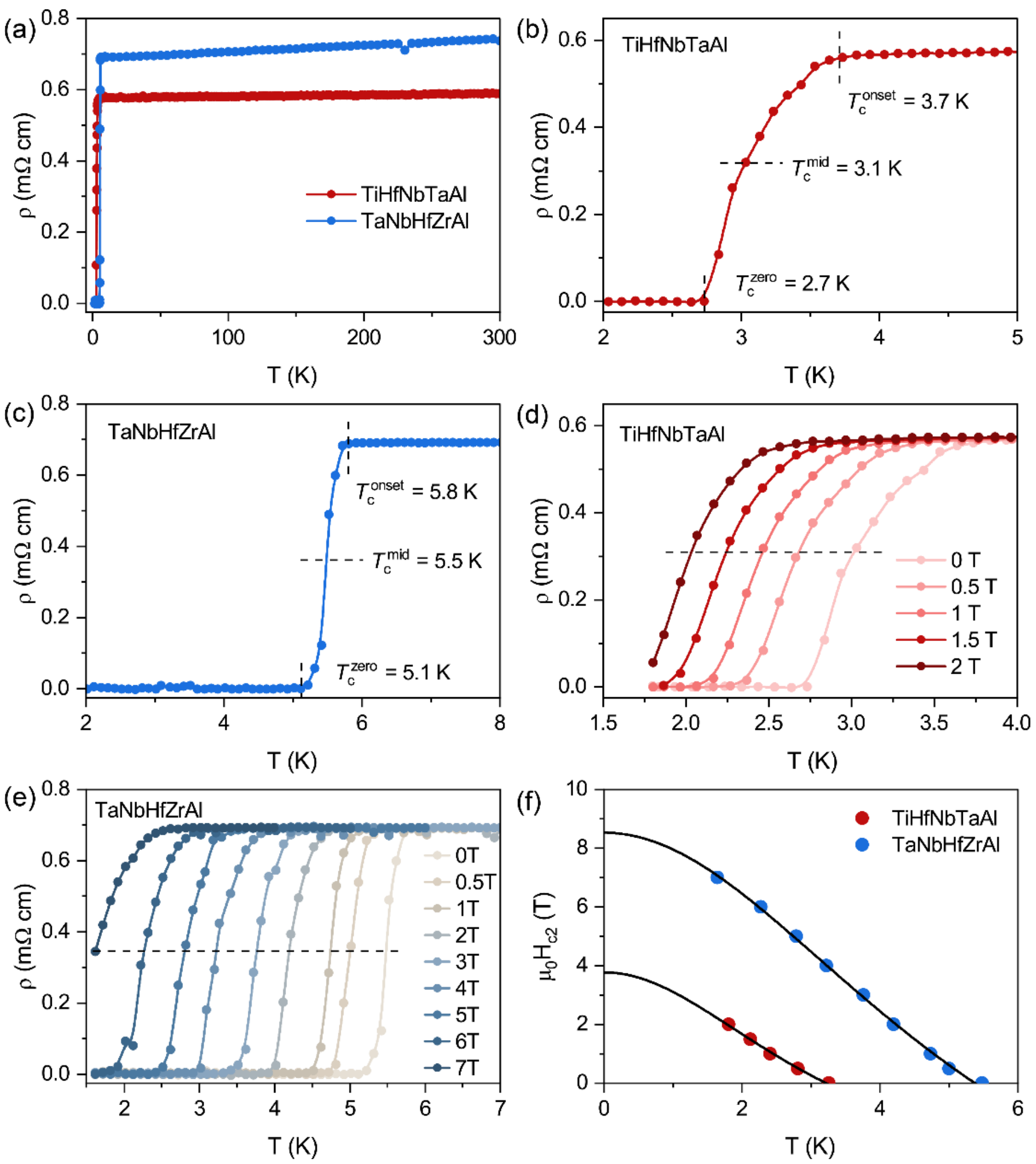

**Figure 2.** (a) The temperature dependence of electrical resistivity for the TiHfNbTaAl and TaNbHfZrAl from 1.8 to 300 K. The detailed superconducting transitions of (b) TiHfNbTaAl and (c) TaNbHfZrAl. The resistivity data for the (d) TiHfNbTaAl and (e) TaNbHfZrAl under applied magnetic fields. (f) The temperature-dependent upper critical field fitting by the GL model for TiHfNbTaAl and TaNbHfZrAl.

We can also demonstrate the superconductivity of these two Al-based HEAs by measurements of the volume magnetic susceptibility. Using volume magnetization $M_v$, we defined the formula $\chi_V = \frac{Mv}{H}$ as the magnetic susceptibility. The magnetization data obtained in the zero-field-cooling (ZFC) model with a 3 mT applied field are shown in Figure 3(a). It shows a strong diamagnetic signal at the $T_c$ = 3.2 K for TiHfNbTaAl and $T_c$ = 5.4 K for TaNbHfZrAl, confirming the emergence of superconductivity. The demagnetization factor N, derived from the M(H) curves, was used to revise the experimental data. The value of the magnetization curve drops to -1, indicating the 100% diamagnetic volume fraction, certifying that these two Al-based HEAs are bulk superconductors.

Figure S2 illustrates the M(H) curves, which depict the magnetization against the applied magnetic field, across various temperatures below the $T_c$ for a more detailed discussion of the characterization. The $M_v$-$M_{fit}$ curves in Figure S2(b) of TiHfNbTaAl and TaNbHfZrAl (Figure S2(d)) were created by linear fitting the low-field magnetization data ($M_{fit}$). The initial point where the slope deviates from linearity is used to estimate the value of the lower critical field $\mu_0 H_{c1}^*$. Figure 3(b) displays all the $\mu_0 H_{c1}^*$ values along with their respective temperatures for TiHfNbTaAl and TaNbHfZrAl. The values of $\mu_0 H_{c1}^*(T)$ are well-fitted through the GL formula $\mu_0 H_{c1}^*(T) = \mu_0 H_{c1}^*(0)(1-(T/T_c)^2)$, where $\mu_0 H_{c1}^*(0)$ represents the lower critical field at 0 K. The GL formula accurately fits the experimental data and gives $\mu_0 H_{c1}^*(0)$ = 17.60 mT for TiHfNbTaAl and $\mu_0 H_{c1}^*(0)$ = 33.94 mT for TaNbHfZrAl. Besides, when revised by the demagnetization factor $N$, the $\mu_0 H_{c1}(0)$ was calculated via $\mu_0 H_{c1}(0) = \mu_0 H_{c1}^*(0)/(1-N)$. The estimated $\mu_0 H_{c1}(0)$ = 20.9(6) mT for TiHfNbTaAl and $\mu_0 H_{c1}(0)$ = 200.4(9) mT TaNbHfZrAl HEA.

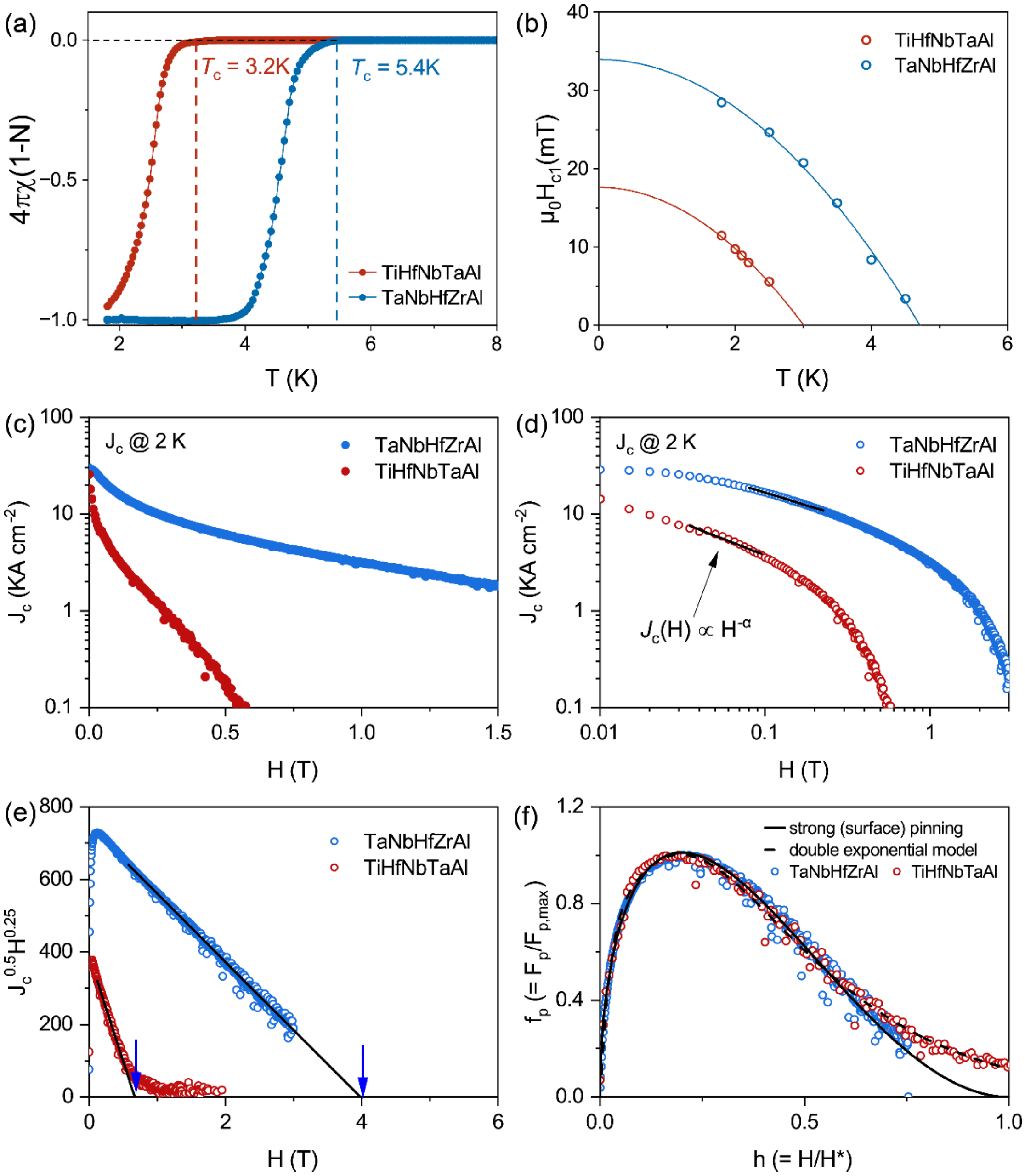

Figure 3. (a) Temperature-dependent ZFC magnetization of TiHfNbTaAl and TaNbHfZrAl HEAs. (b) The temperature-dependent lower critical field $\mu_0 H^*_{c1}$ fitting with the GL model. (c) Semi-log plot of magnetic field dependence of the critical current density Jc of TiHfNbTaAl and TaNbHfZrAl HEAs at 2 K. (d) Log-log plot of magnetic field dependence of critical current density $J_c$ at 2 K for TiHfNbTaAl and TaNbHfZrAl. The solid line is the power-law field-dependent critical current density $J_c(H) \propto H^{-\alpha}$. (e) The Kramer plots, represented by solid lines, are used to identify the irreversible fields $H^*$, as shown by the arrow. (f) Normalized flux pinning force density, $f_p$, with the reduced field h of the TiHfNbTaAl and TaNbHfZrAl HEAs at 2 K.

Figure S3 illustrates the field dependence of isothermal magnetic hysteresis M(H) for TiHfNbTaAl and TaNbHfZrAl at 2 K. The M(H) curves show the characteristic behavior of type-II superconductors. The Bean model is used to estimate the critical current density ($J_c$) through the M(H) curves, with the formula $J_c = 20\Delta M/[a(1-a/3b)]$, where ΔM represents the vertical width of the M(H) loop [24]. Constants a and b are the sample width and length, where a < b. The semi-logarithmic plot of the isothermal $J_c(H)$ with a magnetic field at 2 K for TiHfNbTaAl and TaNbHfZrAl HEAs is shown in Figure 3(c). The $J_c$ of the TiHfNbTaAl and TaNbHfZrAl at 2 K are about 25604 A $cm^{-2}$ and 29754 A $cm^{-2}$, respectively. In comparison, for the typical NbTi binary alloy superconductor, the $J_c$ is about $10^5$ A $cm^{-2}$ at 2 K [25-26].

Table 1. Superconducting parameters of typical MEA/HEA superconductors [16-17, 21].

| **Parameter** | **TiHfNbTaAl** | **TaNbHfZrAl** | **TaNbHfZr** | **TiHfNbTa** | **$Ta_{1/6}Nb_{2/6}Hf_{1/6}Zr_{1/6}Ti_{1/6}$** |
|---|---|---|---|---|---|
| **$T_c$ (K)** | 3.1 | 5.5 | 8.1 | 6.75 | 7.85 |
| **$\mu_0H_{c1}$ (mT)** | 20.9(6) | 200.4(9) | / | 45.8 | 23 |
| **$\mu_0H_{c2}$ (T)** | 3.7(7) | 8.5(3) | 16.3 | 10.46 | 12.05 |
| **$\gamma$ (mJ $mol^{-1}$ $K^{-2}$)** | 3.86 | 3.21 | / | 4.70 | 7.451 |
| **$\beta$ (mJ $mol^{-1}$ $K^{-4}$)** | 0.16 | 0.19 | / | 0.242(8) | 0.274 |
| **$\Delta C_{el}/\gamma T_c$** | 1.54 | 2.31 | / | 2.88 | 2.22 |
| **$2\Delta_0/K_BT_c$** | | 4.48 | / | 5.02 | 4.72 |
| **$\Theta_D$** | 229.7(3) | 216.9(3) | / | 199.9 | 192.3 |
| **$\lambda_{ep}$** | 0.6(1) | 0.7(4) | / | 0.83 | 0.9 |
| **VEC** | 4.2 | 4.2 | 4.5 | 4.5 | 4.5 |

The $J_c(H)$ curves were plotted using a double-logarithmic scale, as depicted in Figure 3(d), to comprehend the pinning mechanism of the TiHfNbTaAl and TaNbHfZrAl HEA. The critical exponent α in the relation $J_c(H) \propto H^{-\alpha}$ was found to be 0.74 and 0.67 for TiHfNbTaAl and TaNbHfZrAl HEA at 2 K, which is close to the value of $Ta_{1/6}Nb_{2/6}Hf_{1/6}Zr_{1/6}Ti_{1/6}$ [16]. The empirical irreversible field $H^*$ was estimated using the Kramer plot $J_c^{0.5}H^{0.25} \propto (H^*-H)$, as depicted in Figure 3(e) [27-29]. The $H^*$ is estimated as 0.71 T and 4.8 T of TiHfNbTaAl and TaNbHfZrAl, respectively. Figure 3(f) illustrates the reduced field h (= $H/H^*$) dependence of normalized flux pinning force density $f_p$ (= $F_p/F_{p,max}$), where $F_p$ represents the flux pinning force density, and its highest value is represented by $F_{p,max}$. We utilized the empirical irreversible field $H^*$ from Figure 3(e) to achieve a reduced field h. The $f_p(h)$ curves are effectively scaled with the double exponential model using $f_p(h) = J_1\exp(-B_1h)h + J_2\exp(-B_2h)h$ [30], and the strong (surface) pinning model of Dew-Huges, $f_p(h) \propto h^{0.5}(1-h)^2$ [31]. These pinning mechanisms pertain to a two-gap superconductor like $MgB_2$ [32-34]. The two-gap structure of these two Al-based HEA compounds is uncertain, but we can infer the existence of two pinning mechanisms.

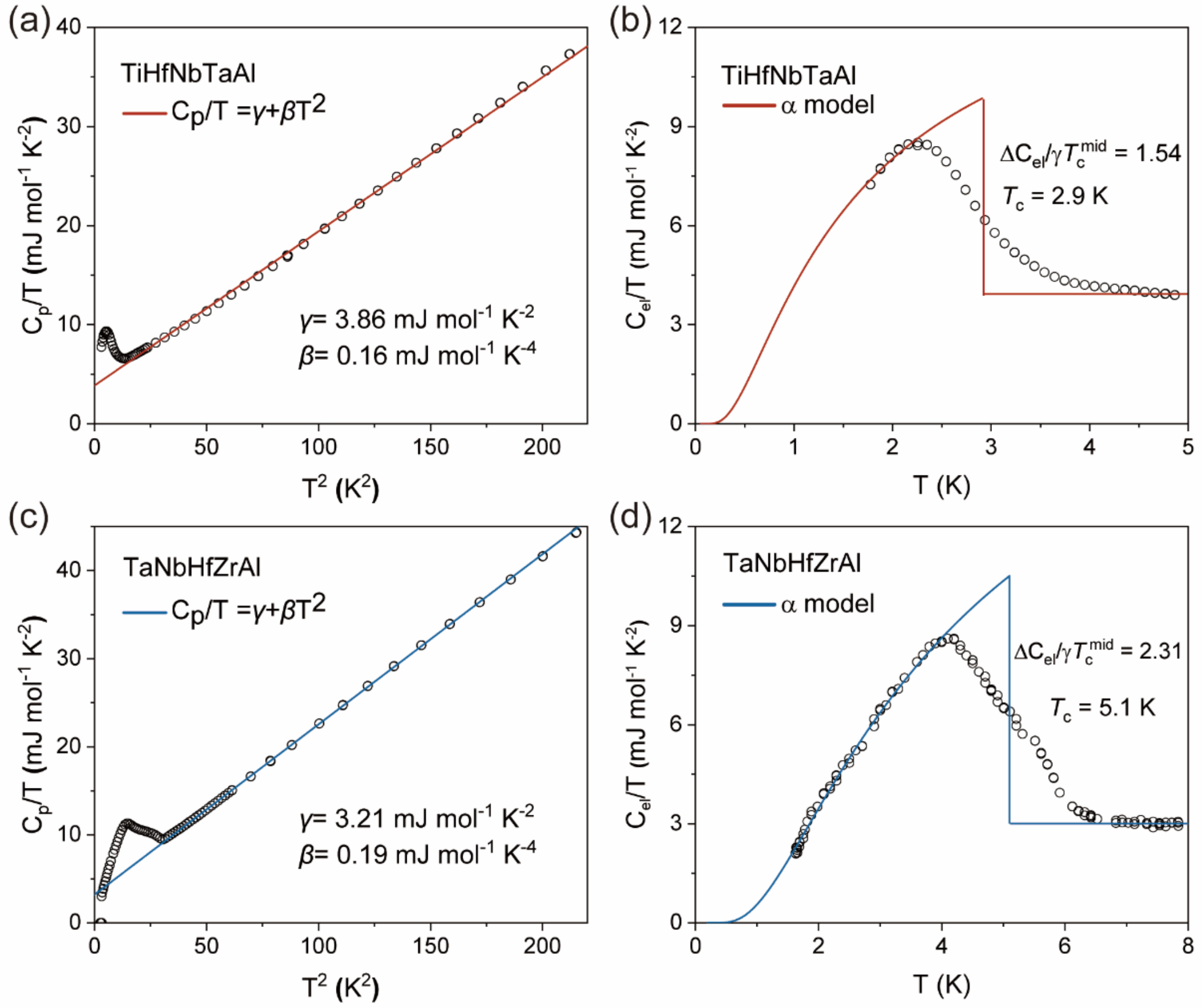

**Figure 4.** The $C_p/T$ curve of (a) TiHfNbTaAl and (c) TaNbHfZrAl HEAs, which utilizes the Debye model $C_p/T = \gamma + \beta T^2$. Electronic part of the heat capacity divided by temperature ($C_{el}/T$) curves of (b) TiHfNbTaAl and (d) TaNbHfZrAl HEAs.

We also measured the specific heat $C_p(T)$ to better understand the thermodynamic properties of these two Al-based HEAs. A sharp jump in Figure 4(a) and Figure 4(c) evidences the bulk superconductivity. The $T_c$ was found as 2.9 K for TiHfNbTaAl and 5.1 K for TaNbHfZrAl, consistent with the magnetic susceptibility and resistivity data. The collected data of the normal state can be fitted by the formula $C_p/T = \gamma + \beta T^2$, where $\beta T^2$ is the contribution of the specific heat of phonons, and $\gamma$ reflects the Sommerfeld constant of the normal state. By fitting the experimental data to the above formula, the Sommerfeld parameter was given as 3.86 mJ mol$^{-1}$ K$^{-2}$ for TiHfNbTaAl and 3.21 mJ mol$^{-1}$ K$^{-2}$ for TaNbHfZrAl. The estimation of $\beta$, based on the slope of each fitted line, was 0.16 mJ mol$^{-1}$ K$^{-4}$ for TiHfNbTaAl and 0.19 mJ mol$^{-1}$ K$^{-4}$ for TaNbHfZrAl. Having the $\beta$ value, we calculated the Debye temperature $\Theta_D$ by the formula: $\Theta_D = (12\pi^4 nR/5\beta)^{1/3}$, where R is the gas constant, and n is the number of atoms in the formula unit. The values of $\Theta_D$ are 229.7(3) K for TiHfNbTaAl and 216.9(3) K for TaNbHfZrAl.

Based on the $T_c$ value and Debye temperature, we can calculate the electron-phonon coupling constant $\lambda_{ep}$ by using the inverted McMillan equation: $\lambda_{ep}=\frac{1.04+\mu^* \ln\left(\frac{\Theta_D}{1.45T_c}\right)}{(1-0.62\mu^*)\ln\left(\frac{\Theta_D}{1.45T_c}\right)-1.04}$, where the Coulomb pseudo-potential $\mu^*$ has a typical value 0.13 [14, 35-36]. The superconducting constant $\lambda_{ep}$ = 0.6(1) for TiHfNbTaAl and $\lambda_{ep}$ = 0.7(4) for TaNbHfZrAl. We also plotted the curve of specific heat of electron ($C_{el}$) for TiHfNbTaAl and TaNbHfZrAl in Figure 4(b) and Figure 4(d) to understand the behavior of the electrons, where $C_{el}$ can be obtained by the formula $C_{el} = C_p - \beta T^3$. We found a clear $C_{el}$ jump in the specific heat of electrons. We fitted the data using the α-model [37]. The conventional *s*-wave gap function effectively describes the experimental data, indicating that these two Al-based HEAs are completely gap superconductors. With the Sommerfeld coefficient, another important superconducting parameter, the specific heat jump at $T_c$ ($\Delta C_{el}/\gamma T_c$), can be determined. The normalized specific heat jump is calculated as 1.54 and 2.31 for TiHfNbTaAl and TaNbHfZrAl, respectively. For the TaNbHfZrAl HEA, the calculated value (2.31) significantly exceeds the expected value of 1.43 for the BCS weak coupling limit, indicating strongly coupled electrons are involved in the superconductivity in this Al-based HEA. The angular independent gap function equation is $\Delta(T) = \alpha/\alpha_{BCS}\Delta_{BCS}(T)$, where $\alpha_{BCS} = 1.76$ is the weak-coupling gap ratio. From Figure 4(d), we are able to obtain the superconducting gap value of 4.48, far exceeding the theoretical value of 3.52 of weak coupling BCS, proving that it is a strongly coupled superconductor. Such a large value of $\Delta C_{el}/\gamma T_c$ was also observed in TiHfNbTa MEA and $Ta_{1/6}Nb_{2/6}Hf_{1/6}Zr_{1/6}Ti_{1/6}$ HEA [16-17]. Comparisons to other strongly coupled systems are made in Table 1.

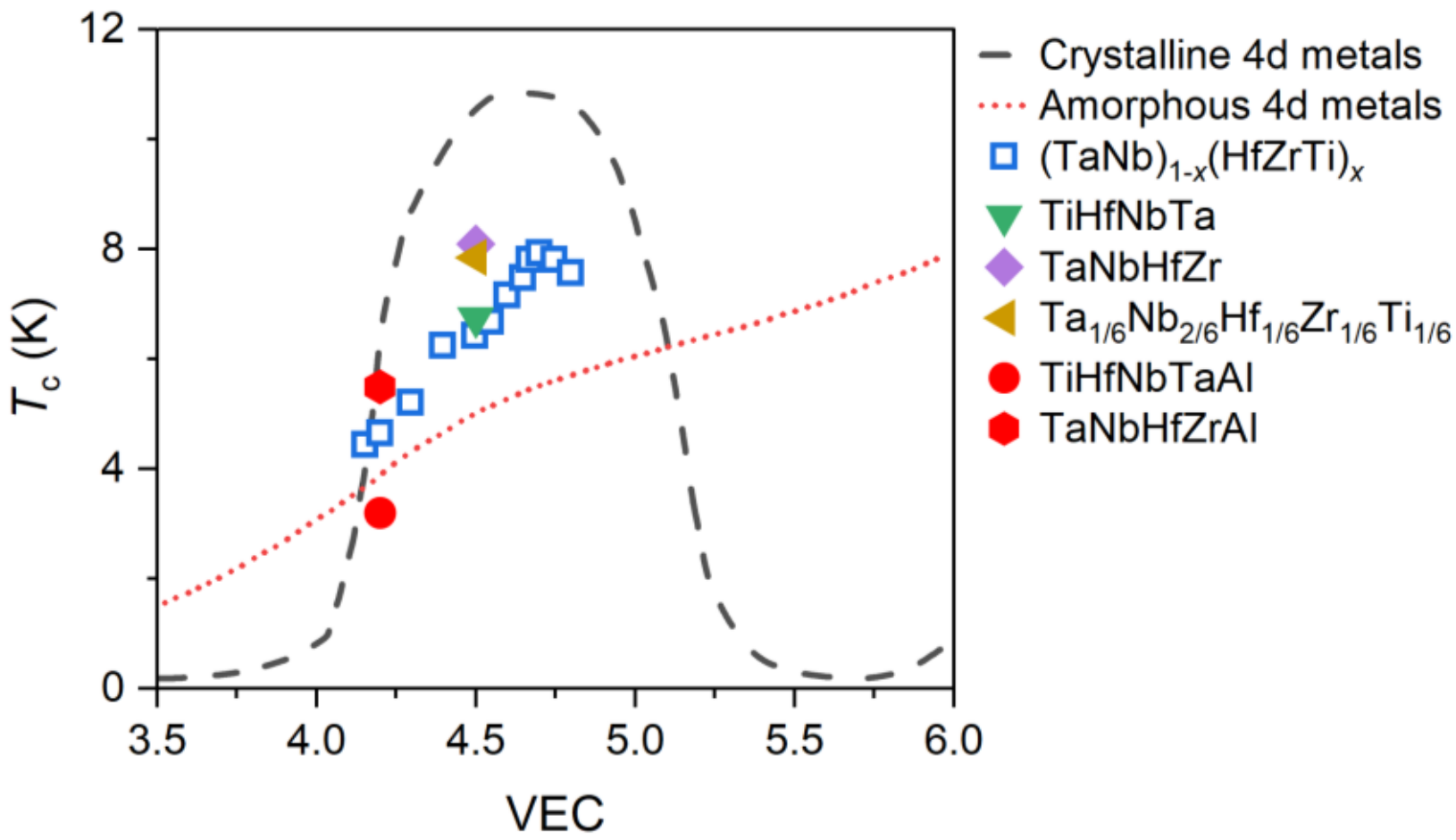

**Figure 5.** The $T_c$ with VEC of TiHfNbTa- and TaNbHfZr-based superconductors [16-17, 21, 38].

To ascertain the electronic properties, including the effective mass ($m^*$), Fermi velocity ($V_F$), and mean free path ($l$), we employ the calculated parameters such as the charge carrier density ($n$), derived from the Hall measurements (see Figure S4) and Sommerfeld coefficient ($\gamma$), obtained from transport measurements. The effective mass $m^*$ was estimated as 2.4(1) me for TiHfNbTaAl and 2.3(1) me for TaNbHfZrAl by the formula $\gamma = (\frac{\pi}{3})^{2/3} \frac{k_B^2 m^* n^{1/3}}{\hbar^2}$, where $\hbar$ is the reduced Planck constant. The $V_F$ is calculated by substituting the values of $n$ and $m^*$ into the formula: $n = \frac{1}{3\pi^2}(\frac{m^* V_F}{\hbar})^3$. We get the Fermi velocity VF = 6.6(4) × 106 m/s for TiHfNbTaAl and 5.9(2) × 106 m/s for TaNbHfZrAl. The mean free path is calculated form the expression: $l = \frac{3\pi^2\hbar^3}{e^2\rho_0 m^{*3} V_F^2}$. The mean free path is calculated as 1.1(3) Å for TiHfNbTaAl and 1.2(8) Å for TaNbHfZrAl, which is relatively low and agrees with other HEA superconductors [38-40]. The high amount of disorder by five distinct element in the crystal structure of TiHfNbTaAl and TaNbHfZrAl causes a low value of the mean free path. According to the BCS theory, the coherence length is calculated as 295 nm for TiHfNbTaAl and 148 nm through the expression: $\xi_0 = \frac{0.18\hbar V_F}{k_B T_c}$. For these two Al-based HEA superconductors, the $\xi_0 > l$, classifying them as a dirty limit superconductor.

Figure 5 illustrates the $T_c$ with the valence electron count (VEC) of the TiHfNbTaAl and TaNbHfZrAl HEAs. For comparison, the compounds in Table 1, the observed trend lines of the $T_c$ for amorphous 4$d$ metals and crystalline 4$d$ metals are illustrated as well [41]. The pattern of $T_c$s for crystalline transition metals is called Matthias rule [42], which ascends monotonically with the increase of VEC until it reaches its maximum at a VEC range of 4.5-4.7. The $T_c$ and VEC relationship of $(TaNb)_{1-x}(HfZrTi)_x$ roughly follows Matthias' rule. In BCC-type HEA superconductors, the $T_c$ is influenced by the VEC and the element composition, particularly the Nb content [20]. The TiHfNbTaAl and TaNbHfZrAl have a VEC of 4.2, which is lower than the 4.5 of TiHfNbTa and TaNbHfZr. And the Nb content has also decreased. Therefore, compared to TiHfNbTa and TaNbHfZr, TiHfNbTaAl and TaNbHfZrAl have lower $T_c$. Furthermore, the addition of 20% Al element in TiHfNbTa and TaNbHfZr causes $T_c$ to decrease rapidly. Compared to the trend line of $(TaNb)_{1-x}(HfZrTi)_x$ HEAs, it becomes evident that the incorporation of Al causes a more crystallinelike $T_c$ dependence, which similar to $(TaNb)_{0.67}(HfZrTi)_{0.33}Al_x$ [43]. These two Al-based HEA superconductors provide a platform for investigating the superconductivity between amorphous and crystalline metals.

## IV. Conclusion

In summary, we report two new Al-based HEA superconductors, TiHfNbTaAl and TaNbHfZrAl, which both exhibit a BCC-type structure within the space group $Im\bar{3}m$. Temperature dependence of magnetic susceptibility, specific heat, and electrical resistivity data confirmed bulk superconductivity with $T_c$ = 3.2 K for TiHfNbTaAl and $T_c$ = 5.4 K for TaNbHfZrAl. The derived superconducting parameters indicate that these two Al-based HEAs are basically type-II BCS superconductors. The M(H) curves revealed that the flux pinning mechanism aligned with the double exponential model from the Dew-Huges formula, including strong (surface) pinning, suggesting that there were two types of pinning mechanisms in TiHfNbTaAl and TaNbHfZrAl superconductors. We find that the $T_c$ is influenced by the VEC and the element composition. And the incorporation of Al in high disorder HEA superconductors causes a more crystallinelike $T_c$ dependence. The EHAs are an interesting model system for the investigation of structure-property relations in intermetallic superconductors. Furthermore, the specific heat jump $\Delta C_{el}/\gamma T_c$ of the TaNbHfZrAl sample reaches 2.31, where the superconducting gap value of 4.48, indicating it is a strongly coupled superconductor. Thus, the TaNbHfZrAl HEA serves as a novel material platform for exploring the interaction of *s*-wave superconductivity with strong electron-phonon coupling.

**Acknowledgments**

This work is supported by the Guangdong Major Project of Basic Research (2025B0303000004), Natural Science Foundation of China (No. 12274471, 12404165), Guangdong Provincial Science and technology Program Project - International Science and technology Cooperation Field (2025A0505020045), Guangzhou Science and Technology Programme (No. 2024A04J6415), Guangdong Basic and Applied Basic Research Foundation (No. 2025A1515010311), the Open Research Fund of State Key Laboratory of Quantum Functional Materials (NO. QFM2025KF004), the State Key Laboratory of Optoelectronic Materials and Technologies (Sun Yat-Sen University, No. OEMT-2024-ZRC-02), and Key Laboratory of Magnetoelectric Physics and Devices of Guangdong Province (Grant No. 2022B1212010008), and Research Center for Magnetoelectric Physics of Guangdong Province (2024B0303390001).

# Supporting Information

# Superconductivity in Al-based high-entropy alloys TiHfNbTaAl and TaNbHfZrAl

Junjin Huang[a,#], Wenbo Sun[a,#], Longfu Li[a], Shuangyue Wang[a], Jingjun Qin[a], Rui Chen[a], Zaichen Xiang[a], Yucheng Li[a], Lingyong Zeng [a,*], Huixia Luo [a,*]

[a] School of Materials Science and Engineering, State Key Laboratory of Optoelectronic Materials and Technologies, Guang-dong Provincial Key Laboratory of Magnetoelectric Physics and Devices, Key Lab of Polymer Composite & Functional Materials, Sun Yat-sen University, Guangzhou 510006, China

# These authors contributed equally to this work

*Corresponding author:

Dr. Lingyong Zeng (zengly25@mail3.sysu.edu.cn)

Prof. Huixia Luo (luohx7@mail.sysu.edu.cn)

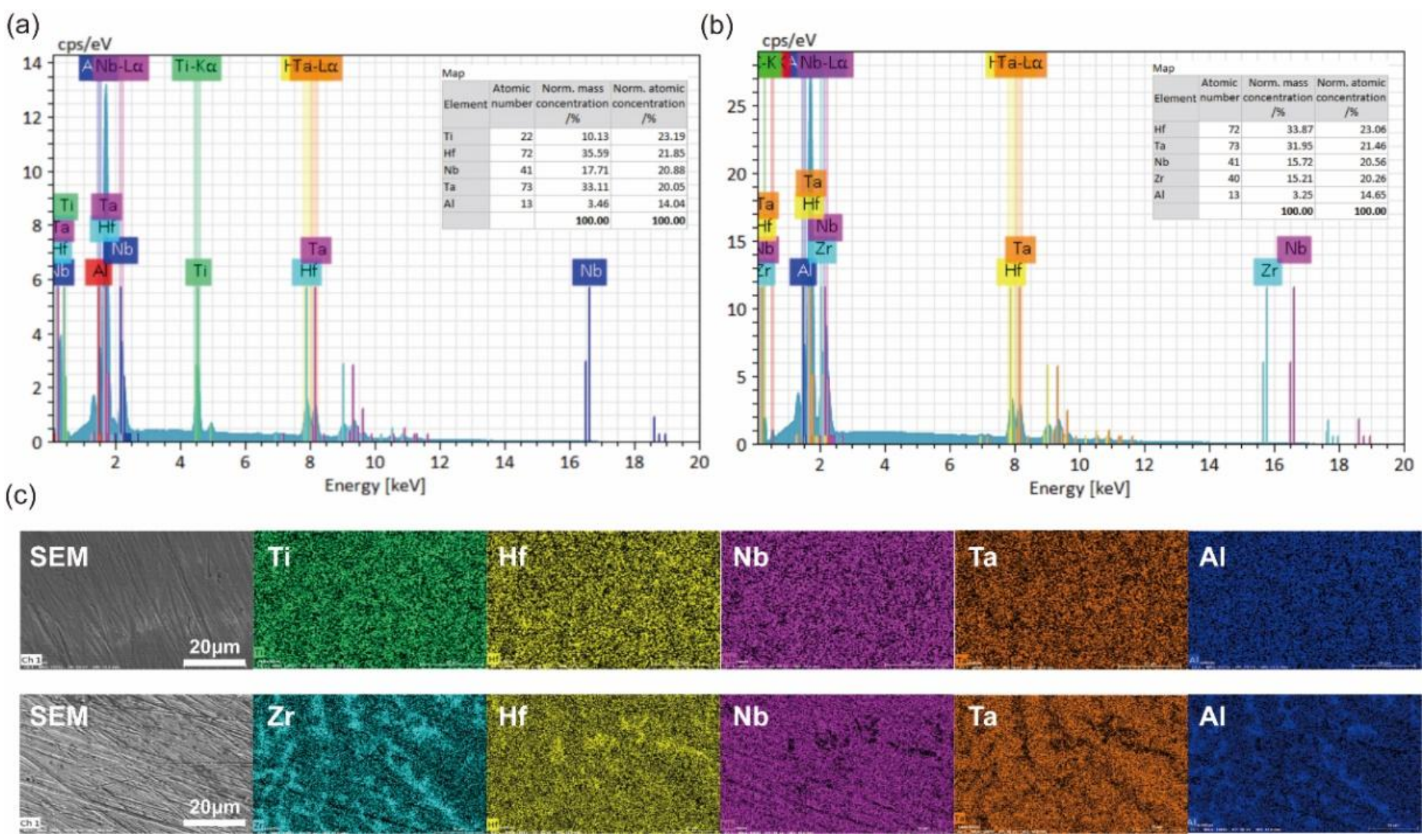

Map

| Element | Atomic number | Norm. mass concentration /% | Norm. atomic concentration /% |
|---|---|---|---|
| Ti | 22 | 10.13 | 23.19 |
| Hf | 72 | 35.59 | 21.85 |
| Nb | 41 | 17.71 | 20.88 |
| Ta | 73 | 33.11 | 20.05 |
| Al | 13 | 3.46 | 14.04 |
| | | **100.00** | **100.00** |

Map

| Element | Atomic number | Norm. mass concentration /% | Norm. atomic concentration /% |
|---|---|---|---|
| Hf | 72 | 33.87 | 23.06 |
| Ta | 73 | 31.95 | 21.46 |
| Nb | 41 | 15.72 | 20.56 |
| Zr | 40 | 15.21 | 20.26 |
| Al | 13 | 3.25 | 14.65 |
| | | **100.00** | **100.00** |

**Figure S1.** EDX spectrum of (a) TiHfNbTaAl and (b) TaNbHfZrAl, respectively. The inset shows the element ratios of these two HEAs. (c) Scanning electron microscope (SEM) images and their elemental maps for TiHfNbTaAl and TaNbHfZrAl HEAs.

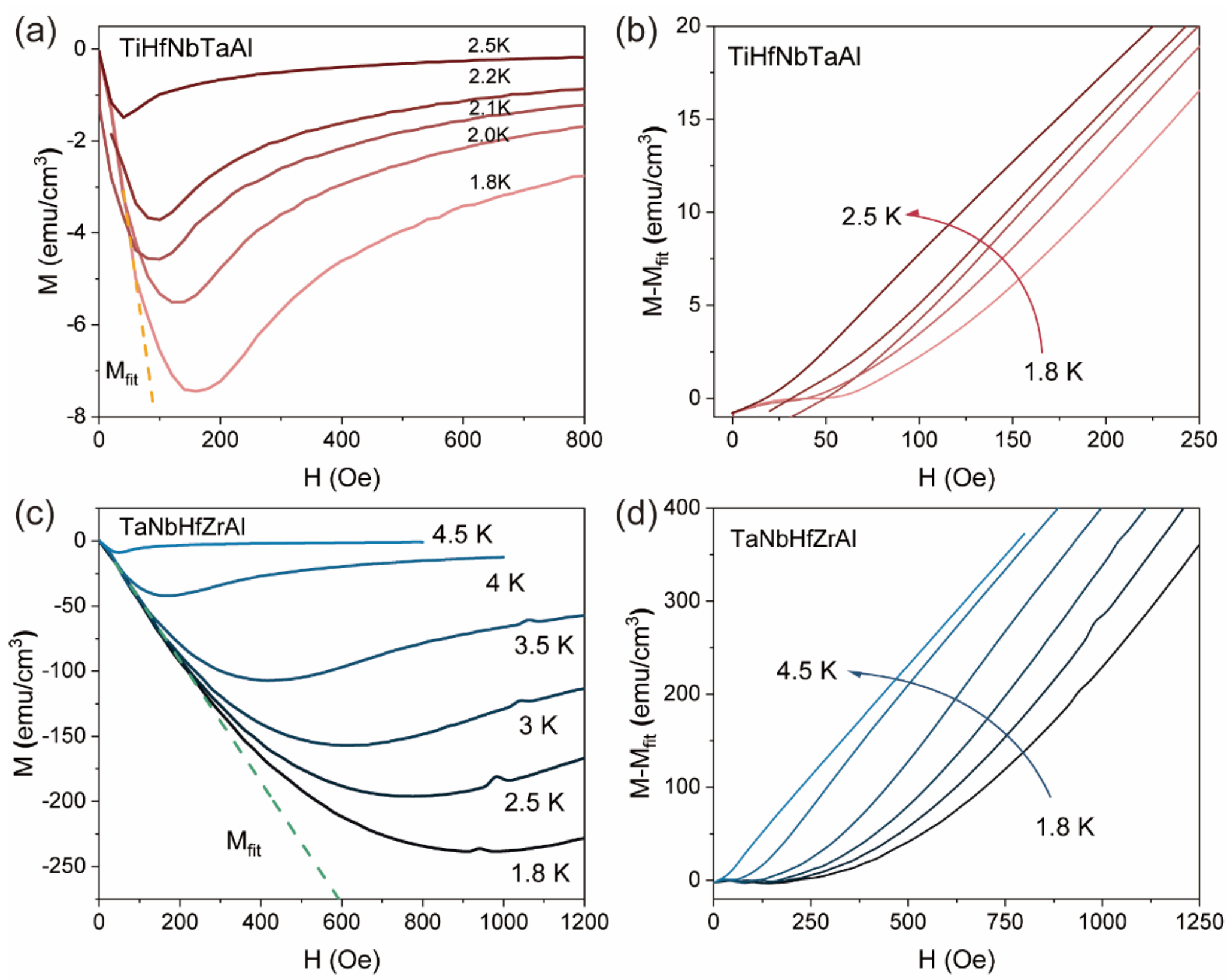

**Figure S2.** The isothermal magnetization curves under different temperatures of (a) TiHfNbTaAl and (c) TaNbHfZrAl, respectively. The difference between M and $M_{fit}$ under different temperatures of (b) TiHfNbTaAl and (d) TaNbHfZrAl, respectively.

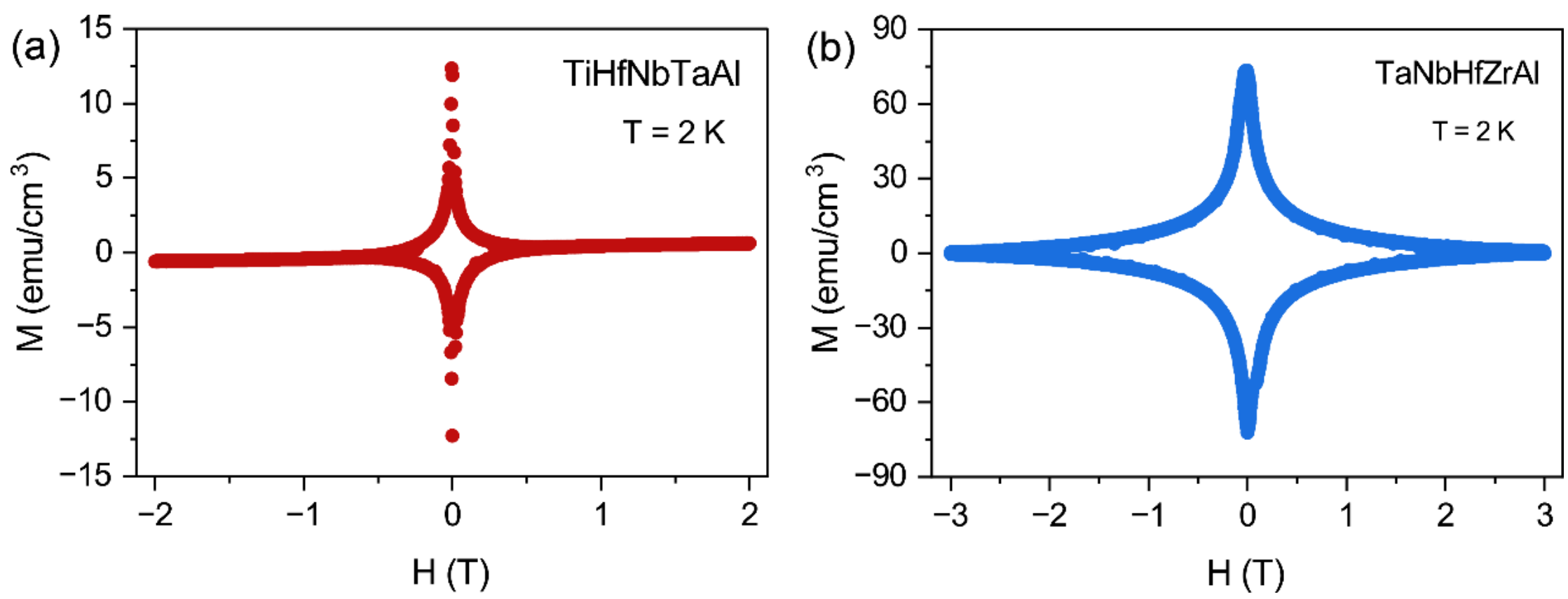

**Figure S3.** Isothermal magnetic hysteresis M(H) loops at 2 K for (a) TiHfNbTaAl and (b) TaNbHfZrAl, respectively.

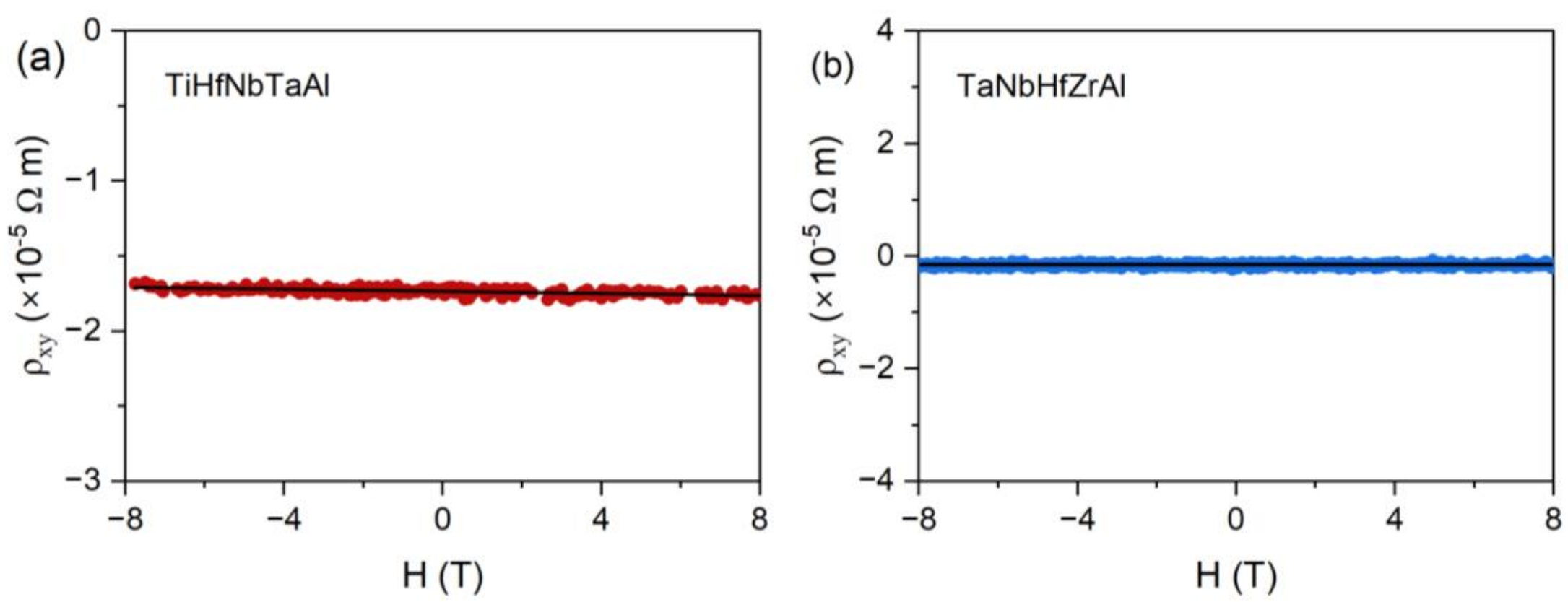

Figure S4. Hall resistivity ($\rho_{xy}$) at 10 K under a magnetic field of ±8T.